\documentclass[referee]{aa}
\usepackage{graphics}

\newcommand{\be}{\begin{equation}}
\newcommand{\ee}{\end{equation}}
\begin{document}


\title{A variability analysis of low-latitude unidentified
gamma-ray sources\thanks{Tables 5-8 are only available in
electronic form at the CDS via anonymous ftp to
cdsarc.u-strasbg.fr (130.79.128.5) or via
http://cdsweb.u-strasbg.fr/cgi-bin/qcat?J/A+A/(vol)/(page) }}

\author{
Diego F. Torres\inst{1,}\thanks{dtorres@venus.fisica.unlp.edu.ar},
Gustavo E. Romero\inst{1,}\thanks{Member of CONICET}, J. A.
Combi\inst{1,\star\star}, P. Benaglia \inst{1,\star\star}, Heinz
Andernach\inst{2}, and Brian Punsly\inst{3}}

\offprints{D. F. Torres}

\institute{Instituto Argentino de Radioastronom\'{\i}a, C.C.5,
(1894) Villa Elisa, Buenos Aires, Argentina \and Depto. de
Astronom\'{\i}a, Univ. Guanajuato, Apartado Postal 144,
Guanajuato, C.P. 36000, GTO, Mexico \and 4014 Emerald Street No.
116, Torrance, CA 90503, USA}

\date{Received 4 August 2000/Accepted 26 January 2001}

\titlerunning{Unidentified gamma-ray sources}
\authorrunning{D. F. Torres et al.}

\abstract{We present a study of 40 low-latitude unidentified 3EG
gamma-ray sources which were found to be not positionally
coincident with any known class of potential gamma-ray emitters in
the Galaxy (Romero, Benaglia \& Torres, 1999). We have performed a
variability analysis which reveals that many of these 40 sources
are variable. These sources have, in addition, a steep mean value
of the gamma-ray spectral index, $<\Gamma> = 2.41 \pm 0.2$, which,
combined with the high level of variability, seems to rule out a
pulsar origin. The positional coincidences with uncatalogued
candidates to supernova remnants were also studied. Only 7 sources
in the sample are spatially coincident with these candidates, a
result that is shown to be consistent with the expected level of
pure chance association. A complementary  search for weak radio
counterparts was also conducted and the results are presented as
an extensive table containing all significant point-like radio
sources within the 40 EGRET fields. We argue that in order to
produce the high variability, steep gamma-ray spectra, and absence
of strong radio counterparts observed in some of the gamma-ray
sources of our sample, a new class of objects should be
postulated, and we analyze a viable candidate.
\keywords{gamma-rays: observations -- gamma-rays: theory -- ISM:
supernova remnants -- black holes physics} } \maketitle


\section{Introduction}

The main results of the very successful Energetic Gamma Ray
Telescope (EGRET) of the Compton Gamma Ray Observatory (CGRO) are
contained in the Third EGRET (3EG) catalog (Hartman et al. 1999),
which includes observations carried out between 22 April 1991 and
3 October 1995, and lists 271 point sources. Of these sources, 170
have no conclusive counterparts at lower frequencies. In the
latitude range $|b| \leq 10\degr$, there are 81 unidentified
sources, a number which doubles that reported in the previous 2EG
catalog and its supplement (Thompson et al. 1995, 1996), and
clearly confirms the existence of a population of gamma-ray
emitters in the Galaxy.

The nature of this population of low-latitude gamma-ray sources is
not clear. Several types of objects have been proposed as possible
counterparts in the past: supernova remnants (e.g. Sturner \&
Dermer 1995, Esposito et al. 1996, Combi et al. 1998a), massive
stars with strong winds (e.g. Montmerle 1979, Romero et al. 1999,
Benaglia et al. 2000), young pulsars (e.g. Yadigaroglu \& Romani
1997), and compact clouds in star-forming regions (e.g. Cass\'e \&
Paul 1980). Gehrels et al. (2000) have recently shown that there
is a population of sources associated with the Gould belt, as it
was originally suggested by Grenier (1995). This suggests that
young stellar objects play an important role in the generation of
the observed emission.

In a recent paper (Romero, Benaglia \& Torres 1999, hereafter
Paper I), some of us presented a study of the level of positional
correlation between gamma-ray sources in the 3EG catalog and a
variety of galactic objects. A statistically very significant
correlation was found in the case of SNRs and OB star
associations. In that work it was also found that a group of 42
low-latitude 3EG sources do not present a likely counterpart. Two
of these sources were finally discovered to be artifacts related
to the high intensity of the Vela pulsar (Hartman 2000). Very
recently, Zhang et al. (2000) proposed that most of the sources
spatially coincident with SNRs and OB associations could be
pulsars. They have presented a variability study of the 38 3EG
sources that are coincident with objects of these classes, arguing
that most of them are not variable.

In this work we focus our attention on the group of 40 gamma-ray
sources which are not coincident with any known galactic object
thought to be capable of producing gamma-ray emission. Classical
galactic counterparts are expected to be non-variable on
timescales of a few years and, consequently, we shall discuss the
time behavior of the sources in our sample. The structure of the
paper is as follows. In Section 2 we define the sample. A
variability analysis is presented in Section 3. Section 4 deals
with the possible positional correlation between sources in our
sample and SNR candidates not included in Green's (1998) catalog.
In Section 5 we explore possible weak radio counterparts of the
gamma-ray sources and present an extensive table with all
significant point-like radio sources within the 40 EGRET
positional error boxes that, we expect, will be very helpful for
future research in this field. In Section 6 we discuss the case
for isolated Kerr-Newman black holes as possible generators of
some of the gamma-ray sources, and finally, in Section 7, we draw
our conclusions. Three appendices present detailed tables with
data used in our research.

\section{Sample}

\mbox{}From the original set of 81 (75 if we exclude the six
sources thought to be artifacts produced by the Vela pulsar)
unidentified sources at $|b|\leq 10\degr$, an apparently large
fraction of $\sim $ 50\% (40 sources) remain without any known
likely galactic counterpart. We list these sources in Table~1,
where we give their 3EG name, their galactic coordinates, the
error in their position --assumed to be represented by the 95\%
confidence contours given by the 3EG catalog--, and their gamma
spectral index $\Gamma$ (such that the photon distribution is
given by $N(E) \propto E^{-\Gamma}$). We also present four further
columns related to the variability analysis that will be explained
below. A $\dag$-symbol on the source name indicates a source which
was suggested as possible AGN by Hartman et al. (1999). It is
worth noticing that the mean value of the spectral index is quite
steep for these sources: $2.41 \pm 0.2$, steeper that the steepest
pulsar spectrum known (Thompson et al. 1994). The distribution of
the spectral index is shown in Figure~1. Only three sources out of
40 have $\Gamma < 2$.

\section{Gamma-ray variability}

We shall now assess the possible long term variability (on
time-scale of years) of the sources in our sample. We will adopt
the following criterion. We define a mean weighted value for the
EGRET flux as: \be \left< F \right> = \left[ \sum_{i=1}^{N_{{\rm
vp}}} \frac{F(i)}{\epsilon(i)^2} \right]\times \left[
\sum_{i=1}^{N_{{\rm vp}}} \frac 1{\epsilon(i)^{2}}
\right]^{-1}.\ee Here, $N_{{\rm vp}}$ is the number of viewing
periods for each gamma-ray source. We take into account only
single viewing periods (i.e. periods like 305+, virgo, or the
combined P12, P123, and P1234 were not taken into account). $F(i)$
is the observed flux in the $i^{{\rm th}}$-period, whereas
$\epsilon(i)$ is the corresponding error in the observed flux. We
have taken these data directly from the 3EG catalog. For those
observations in which the significance ($\sqrt{TS}$ in the EGRET
catalog) is greater than 3$\sigma$, we took the error as
$\epsilon(i) = F(i)/\sqrt{TS}$. However, many of the observations
are in fact upper bounds on the flux, with significance below
3$\sigma$. For these ones, we assume both $F(i)$ and $\epsilon(i)$
as half the value of the upper bound. We then define the
fluctuation index $\mu$ as: \be \mu =100\times \sigma_{{\rm
sd}}\times \left< F \right>^{-1} .\ee In this expression,
$\sigma_{{\rm sd}}$ is the standard deviation of the flux
measurements, taking into account the previous considerations.

Results for the 40 sources are given in Table~1, where we show
four columns corresponding to $\left< F \right>,\sigma_{{\rm
sd}},N_{{\rm vp}}$ and $\mu$.

In order to remove as far as possible spurious variability
introduced by the observing system, we computed the fluctuation
index $\mu$ for the confirmed gamma-ray pulsars in the 3EG catalog
(see Cusumano et al. 2000 for the latest identification). We adopt
the physical criterion that pulsars are --i.e. by definition--
non-variable gamma-ray sources. Then, any non-null $\mu$-value for
pulsars is attributed to experimental uncertainty. We then define
an averaged statistical index of variability, $I$, as \be
I=\frac{\mu_{{\rm source}}}{<\mu>_{{\rm pulsars}}}=\frac{\mu_{{\rm
source}}}{26.9}. \ee \mbox{}

Table~2 shows the variability results for the pulsars in the 3EG
catalog, whereas the last column of Table~1 shows the results of
the variability index $I$ for our sample of unidentified 3EG
sources.

The adopted variability criterion is then to consider that any
source for which the $\mu$-value is less than the upper value of
$\mu$ for pulsars ($\mu_{\rm max}=40.5$), is a non-variable
source. At least, we can assure that for this kind of source, and
with the data now at hand, it is not possible to discriminate any
significant long-term variation in its gamma emission. We shall
also consider that sources with $\mu$-values between 40.5 and 65
are dubious cases, and that sources with $\mu>65$ are variable. In
terms of the averaged index $I$, this is equivalent to saying that
variable sources will be those with $I>2.5$, which is 3$\sigma$
away from the statistical variability of pulsars. The value of
1$\sigma$ is naively obtained from the standard deviation of the
$I$-index for pulsars. Since this is a population of just a few
members, this forces us to be careful in assigning to a source the
``variable'' status.

We adopted this criterion, previously used in blazar variability
analysis by some of us (Romero et al. 1994) and applied to 3EG
sources by Zhang et al. (2000), rather than the similar one used
by McLaughlin et al. (1996) and Wallace et al. (2000) in order to
allow a direct comparison with the spurious statistical
variability shown by pulsars. Since the $I$-index establishes how
variable a source is with respect to the pulsar population, it can
be considered as indicative of how reliable we can be about the
possible physical variation in the gamma emission. The combination
of the results of Zhang et al. (2000) with those presented in this
paper yields the variability index for all unidentified
low-latitude sources in the 3EG catalog, see Appendix C. Note,
however, that Zhang et al.'s study differs from the present one in
that they considered PSR B1951+32 for the determination of
$<\mu>_{{\rm pulsars}}$. This pulsar is not in the 3EG catalog.
The fluxes used by Zhang et al. were taken from data reduced using
a different technique. The inclusion of this pulsar makes
$<\mu>_{{\rm pulsars}}=77.7 \pm 50.0$ instead of the much lower
value we obtain. This high value of $<\mu>_{{\rm pulsars}}$ causes
the $I$-index for AGNs to be compatible with a non-variable
population, and it is also the reason why the majority of Zhang et
al.'s sources are non-variable ones. In Appendix C, the values of
$I$ for all sources at low latitudes ($|b| \leq 10^o$) are
computed under our normalization.

A more comprehensive study on the long term variability (more than
one month) of the gamma-ray sources was recently presented by
Tompkins (1999). This work re-analyzed the EGRET data to take into
account not only all sources included in the 3EG catalog, but also
the 145 marginal sources that were detected but not included in
the final official list. The maximum likelihood set of source
fluxes was then computed. From those fluxes, a new statistics
measuring the variability is defined as $\tau=\sigma / \mu$, where
$\sigma$ is the standard deviation of the fluxes and $\mu$ their
average value. The strength of this approach lies in that it takes
into account possible fluctuations from the background and from
neighboring sources, careful sensitivity corrections throughout
EGRET lifetime, and others systematic errors. On the other hand,
our method can be very useful as a first approach: when one takes
a fairly safe assumption as the threshold for variability (such as
the one introduced above, i.e. a source is classified as variable
if $I>2.5$) it yields results that are compatible with Tompkins'.
For instance, if we consider the sources that Tompkins found as
the most variable ones, we find that only two of them are included
in our sample. These have $I=2.96$ and $I=5.33$, respectively. If
we now consider the sources that Tompkins classified as less
variable, we find that two of them are in our sample, with indices
$I=1.00$ and $I=1.86$. The rest of the cases are compatible too,
but less strength should be put on the results, since most of them
are dubious cases in both schemes. We then found that our method,
avoiding most of the criticism of previous studies (Tompkins 1999)
is well-fitted as a discriminator between variable and
non-variable sources using the publicly available EGRET data.

With our adopted criterion, we find that only a low 25.0\% of the
sources in our sample are non-variable, 42.5\% are dubious cases,
and 32.5\% are variable. In this latter group, three sources have
a variability index 5$\sigma$ away from the statistical variations
of pulsars. In Figure~2 we present the histogram for the
variability index $I$ in our sample, along with a Gaussian fit
which presents its peak at $I=$2.0 and a standard deviation equal
to $\sigma_I=$0.7. In order to better evaluate these results, we
have also computed the variability index for the 66 confirmed AGNs
in the 3EG catalog. A table with these results (with the same
columns as in Table~1) is given in Appendix B. Fifty-eight AGNs
appear to be variable sources. We show the distribution of the
variability index for AGNs in Figure~3. The Gaussian fit in this
case peaks at $I=$2.3 and the standard deviation is
$\sigma_I=$0.8. Since the distribution of the unidentified sources
under study is similar to that obtained for AGNs, which is a
well-known variable population, we can conclude that many of our
EGRET sources may actually vary on time-scales of years.

In Fig.~4 we plot the computed value of $I$ versus the observed
value of the spectral index $\Gamma$, for the sources in our
sample. We also show in the same figure the $I$ - $\Gamma$
relation for AGNs. Two solid horizontal lines represent the lower
and upper bound for the statistical variability of the pulsar
population. An horizontal dashed line stands for the value of the
variability indices above which the variability of the sources can
be established (i.e. above 3$\sigma$ level). From these plots, it
appears that more than a single population is present. Pulsars
should be located at the lower left corner of the figure: they
have a hard spectral index and are non-variable. A different group
of sources, with spectral index in the range $\sim 2.1 - \sim 2.6$
and being marginally variable or variable, appears towards the
center of the figure. Most of the AGNs are located in this region
of the plot. Finally, some sources seem to show a steepening in
their spectral index with increasing values of variability. A
Spearman correlation test for the variable sources yields a
probability of about 8\% for this being a pure chance effect.
These sources could then be members of a new population of
galactic objects or, perhaps, they might be AGNs seen through the
disk of the Galaxy.

Since AGNs are isotropically distributed, we can determine a mean
value for the number of AGNs detected by EGRET per square degree:
$N_{{\rm AGN}}=8.9 \times 10^{-4}$ AGN/degree$^2$. At latitudes in
the range $-10\degr \leq b \leq 10\degr$, the number of
unidentified sources that are marginally variable or variable
yields a number density $N_{{\rm unid}}=4.3 \times 10^{-3}$
objects/degree$^2$. This shows that there are five times more
variable objects within the galactic plane than away from it.
Then, we have to conclude that if the distribution of AGNs is
isotropic, we have an excess of variable galactic sources
representing a new class of population, as it was already noticed
in previous studies by McLaughlin et al. (1996), by Mukherjee et
al. (1997), by Tavani et al. (1997), and very recently by Wallace
et al. (2000).\footnote{The latter work focuses only on very short
timescales.}

Considering the sample of the 2EG catalog, \"{O}zel and Thompson
(1996) argued that one half or more of the unidentified EGRET
sources lying at high latitudes are of galactic origin. Nice and
Sawyer (1997) found no evidence that pulsars were the constituents
of this putative galactic population. This agrees with our
results, since we show here that this new class of galactic
objects should resemble, from the point of view of gamma-ray
observations, the AGN population.

It is perhaps appropriate to come back to the discussion of what
is a real source (as opposed to possible noise variations) in this
study. We are considering that the 3EG  catalog is composed of
real sources, disregarding the level of flux detected in each
case. The exception to this rule is the six artifacts related to
the Vela pulsar, which disappear in a map where the Vela emission
is suppressed (Hartman et al. 1999). However, since many
detections in the EGRET catalog are below the $3\sigma$
significance, and the fluxes are compatible with zero, a more
careful treatment is in order. We can consider only those sources
with $\sqrt{TS}
>5\sigma$, which implies a flux above 30 $\times 10^{-8}$ ph
cm$^{-2}$ s$^{-1}$. In that situation we are left only with 16
sources, four of them having $I>2.5$. One of these sources is 3EG
1828+0142, with $I=5.33$, to which we devoted a separate work
elsewhere (Punsly et al. 2000). If the flux threshold is lowered
to $18 \times 10^{-8}$ ph cm$^{-2}$ s$^{-1}$, we are left with 35
sources, and now 11 of them have $I>2.5$. Within this group is 3EG
1735-1500, which has $I=8.86$. It is then apparent that variable
sources exist even when the flux threshold is raised.

The relevance of the previous point lies mainly in the galactic
longitude distribution of the 40 3EG sources of Table~1. In the
left panel of Fig.~5 we show the histogram distribution with
galactic longitude of all sources in our sample; it is clear that
it is peaked towards the galactic center and the galactic
anti-center. Although it is true that the EGRET exposure was
greater in these regions, it appears that the number of sources of
this sample that were detected at these positions is relatively
larger than what results when the whole 3EG catalog is considered.
If we take a more strict limit on the flux, such as given by
taking sources only with $\sqrt{TS}
>5\sigma$, we obtain the distribution shown on the right panel.
That is, a more restrictive limit on the adopted minimum flux will
preferentially delete the anti-center sources, leaving strong
detections towards the galactic center, which is certainly more
compatible with a real galactic population.

It seems to be possible that at least a part of this excess in the
number of sources towards the galactic center could be the
low-latitude tail of the new population found by Gehrels et al.
(2000) at mid-latitudes. This speculation is supported by the fact
that the average spectral index of the sources associated with the
Gould belt is 2.49 $\pm$ 0.04, quite compatible with the indices
of the sources in the sample here considered. However, the sources
in the mid-latitude population are weaker than many of the
variable and strong ones of our group. If stars are responsible
for the weaker sources, as suggested by Gehrels et al. (2000) and
by Benaglia et al. (2000), then a different type of object,
intrinsically much more luminous at gamma-rays, should be behind
the stronger low-latitudes EGRET detections.

A preliminary conclusion of this section could then be stated as
follows: A new type of galactic objects may be needed to explain
the behaviour of some unidentified 3EG sources at low latitudes.
This population should be able to display high luminosities, steep
gamma-ray spectral indices and significant gamma-ray variability.
Fig. 4 suggests that our sample, although mostly of galactic
origin, mimics some observational properties of AGNs.

\section{Uncatalogued supernova remnants}

We consider in this section whether some of the sources in our
sample may be associated with recently proposed candidates to
supernova remnants, not catalogued by Green (1998). Our interest
in this search resides in the fact that young stellar objects like
recently formed black holes and pulsars can still be associated
with the gaseous remnant of the original supernova that created
them.

The diffuse non-thermal emission of the galactic disk, formed by
the interaction of the leptonic component of the cosmic rays with
the galactic magnetic field, is surely veiling many remnants of
low surface brightness. Recent observational studies using
filtering techniques in the analysis of radio data have revealed
several new SNR candidates that are not yet included in the latest
issue of Green's catalog, which was used in the study presented in
Paper I (e.g. Duncan et al. 1995, Combi \& Romero 1998, Combi et
al. 1998b, 1999a, Jonas 1999). In general, these new candidates
are much more extended than those previously known. There are 101
of these weak non-thermal structures detected so far in the
Galaxy. This number significantly extends Green's (1998)
catalogue. The list of these new candidates, compiled from the
papers and studies above mentioned is being published
electronically as an Appendix.

With the aim of finding the positional coincidences between our
sample of 40 3EG unidentified sources and the uncatalogued
candidates to SNRs, we used the code developed in Paper I. We have
found that only 7 gamma-ray sources in our sample are positionally
coincident with non-thermal radio structures. The positional
coincidences thus obtained are shown in Table~3, where we give the
name of the gamma-ray sources, the central position of the SNR
candidates, the distance from the EGRET source to the centre of
the radio structure ($\delta$), and the addition of the radius of
the extended sources and the uncertainty in the position of the
EGRET detections ($\Delta$).

To estimate the statistical significance of these coincidences, we
have numerically simulated a large number of synthetic sets of
EGRET sources using the code described in Paper I. As in that
paper, the simulations were constrained to preserve the original
gradient in the number distribution of unidentified gamma-ray
sources towards the galactic plane. The results of this study show
that the expected number of chance associations is 10.4 $\pm$ 2.7,
quite compatible with, and even bigger than, the actual result. We
conclude, then, that there is no statistical evidence suggesting
that the 3EG sources analyzed in our sample are associated with
known or potential SNRs.

\section{Search for extended radio counterparts in selected sources}

We have especially examined the radio sky towards the best
estimate positions of the unidentified 3EG sources J0241+6103,
J0435+6137, J0628+1847, J1014-5705, J1631-4033, J1735-1500,
J1828+0142, J1928+1733, J2035+4441 and J2206+66. From Table~1, it
can be seen that these sources can be grouped into two sets, at
the extremes of the variability and spectral index ranges. On one
hand, we have six sources with low variability and relatively hard
spectral indices. These sources could, in principle, be associated
with pulsars or with unknown new SNR candidates. Only one of them,
3EG J1631-4033, presented positional superposition with a known
candidate to SNR (see Table~3). There are, on the other hand, four
sources with high variability ($I$-values ranging from 3.3 to 8.8)
and steep spectral index ($\Gamma$ ranging from 2.08 to 3.24).

We used continuum radio data at 1420 and 2400 MHz, from the
surveys by Reich \& Reich (1986), and Duncan et al. (1995), in
order to study the possible presence of weak and extended radio
sources within the EGRET positional error boxes of this set.
Specifically, we searched for yet undetected low-brightness SNRs
that could be associated with the pulsars or the black holes
created in the original supernova explosion. We remove the
background diffuse radiation using the Gaussian filtering
technique described in a recent series of papers by Combi and
coworkers (e.g. Combi \& Romero 1995, 1998; Combi et al. 1998a, b,
1999a,b). The reader is referred to these papers for details.
Three of the fields that were studied presented features
interesting enough to deserve individual description. Below we
briefly comment on two of them: 3EG J0435+6137 and 3EG J1631-40.
The source 3EG J1828+01 is discussed separately in the paper by
Punsly et al. (2000). The remaining fields were empty of extended
radio structures or too confused even after background subtraction
as to provide reliable information.

\subsection{3EG J0435+6137}

In Fig.~5, upper panel, we show the filtered radio image of the
region around 3EG J0435+6137 at 1.42 GHz (with an rms noise of 20
mK and angular resolution of 34$'$) super-imposed to the EGRET
probability contours. Two apparently extended radio sources are
visible in the field. The strongest one, 87GB J0443+6118, is
partially outside the 95\% confidence contour of the gamma-ray
detection. Another source, named 87GB~J0441+6145, is located 20$'$
away from the best estimated position of 3EG~J0435+6137, well
within the 95\% confidence contour.

In Fig.~5, lower panel, we show the central region of the radio
field at 5 GHz and a resolution of 3.5' (Condon et al. 1989).
Several point-sources are visible in the frame. The identification
of the main ones is also indicated in the figure. None of these
sources is known to be extragalactic. One of them, 87GB
J0442+6140, is coincident with the X-ray source 1RXS
J044239.3+61404. Another source, 87GB~J0435+6137 (also known as
TXS 0431+615), is located exactly at the best estimated position
of 3EG J0435+6137. This source has a steep spectral index of $\sim
-1.2$.

We have made Gaussian convolutions of those groups of point
sources that are nearly coincident with the apparent extended
sources seen in the lower resolution map of the upper panel, using
a beamwidth of 34'. We have found that both apparent structures
are likely artifacts resulting from the merging of the several
point-like sources in the region and that, consequently, no
extended non-thermal source can be associated with 3EG~J0435+6137.
The counterpart, if it exists at all, should be one of the point
sources. The most promising candidates seems to be
87GB~J0442+6140, which is also detected at X-rays, and
87GB~J0435+6137, which is the strongest one at 5 GHz.

\subsection{3EG J1631-4033}

Figure~6 shows the filtered radio continuum map of the field
around 3EG J1631-4033 at 2.4 GHz (with an rms noise of 12 mJy
beam$^{-1}$ and an angular resolution of 10.4$'$). The confidence
contours of the likelihood test statistics of the EGRET detection
are also shown as a superposed gray-scale. Four point-like radio
sources can be identified with PMN sources (Griffith \& Wright
1993), namely (listed by increasing Galactic longitude), PMN
J1627-3952 (noted as S1 in Figure~6), PMN J1631-4015 (S2), PMN
J1636-4101 (S3), and PMN J1631-3956 (S4) . The angular separation
between the location of the highest likelihood test statistic for
the gamma-ray source and the position of PMN J1631-4015 is just
$\sim$ 22 arcminutes. The strongest source, located at $(l,b) \sim
(341.9^{\circ}, +4.15^{\circ})$, can be identified with PMN
J1636-4101. We summarize the radio information on the point
sources in Table~4. Any of these sources can be considered as a
potential counterpart. Additional and weaker sources in this
region are listed in Table~5 (see below).

Another interesting feature revealed by the radio map is the
existence of a minimum in the continuum emission towards the
centre of the field at $(l,b) \sim (341.2^{\circ},+4.6^{\circ})$.
This might be the relic of an old explosive event in the ISM in
this region.

\section{Point-like counterparts: systematic search}

We have searched CATS (Verkhodanov et al. 1997), SIMBAD and NED
databases for any possible radio counterparts for the gamma-ray
sources in our sample. We have prepared a table containing the
results of this search. This table (Table~5, which is published
only in electronic form due to its length) lists all significant
point-like radio sources, with flux densities above 30 mJy at
least at one wavelength, found within the 95 \% EGRET confidence
contours of the 40 gamma-ray sources of Table~1.

In Table~5 there are entries for each gamma-ray source listing,
from left to right, radio survey used, source name, right
ascension and declination with the corresponding uncertainties,
observing frequency, flux density and its error, and comments. The
different fields are separated, and within each field (listed in
increasing observing frequency), a different paragraph corresponds
to each radio source. All listed sources (about 600, ranging from
just a few to almost 50 per 3EG field) are located within the
EGRET 95\% confidence contours. The material contained in this
table could be very useful for future research in this field.

\section{Discussion: The black hole hypothesis}

The existence of a number of variable gamma-ray sources with steep
spectra and without strong radio or X-ray counterparts can be
explained postulating a galactic population of magnetized black
holes (Punsly 1998a, b, 1999). In these objects, strong magnetic
fields ($\sim 10^{11}$ G) are supported by an orbiting charged
ring or disk. The black hole itself is also charged and rotating
(Kerr-Newman black hole), but the entire configuration has zero
net charge, in such a way that it is not quickly neutralized by
the accretion of diffuse interstellar matter. The resulting object
has a magnetosphere similar to that presented by pulsars. However,
because of the absence of a solid surface, thermal X-rays are not
produced, as in a neutron star. In addition, there is no accretion
disk capable of generating strong X-ray emission as is expected in
microquasars (Mirabel and Rodriguez 1998). The radiation of the
system, which is non-pulsating due to the alignment of the
rotation and magnetic axes, is produced by synchroton and inverse
Compton mechanism in two relativistic electron-positron jets which
propagate along the rotation axis in opposite directions, as it
occurs in AGNs. Contrary to the case of most AGNs, no redshifted
emission lines should be present and, in contrast to BL Lacs, no
weak nebulosity is expected in the optical band. Punsly et al.
(2000) have recently shown that, when pair annihilation effects
are taken into account in the calculation of the spectral energy
distribution presented by magnetized black holes, a steep
gamma-ray spectrum is produced. Variability naturally results in
the model as the effect of shocks and firehose instabilities which
induce large changes in the Doppler enhancement factor of the
relativistic jets (Punsly et al. 2000).

At high energies, self-Compton losses provide most of the
gamma-ray luminosity, reaching values in the range
$10^{34}-10^{35}$ erg s$^{-1}$ for black holes of a few solar
masses and polar magnetic fields of $\sim10^{11}$ G (see Punsly et
al. 2000 for details). If such an object is relatively close (e.g.
$\sim1$ kpc), it could appear as a typical unidentified EGRET
source with $\Gamma\sim 2.5$ or steeper. Although the spectral
energy distribution is broad, only weak emission at longer
wavelengths is expected. Radio counterparts, for instance, would
present flux densities on the order of a few to a few tens of mJy
at 5 GHz. Interestingly, one would also expect a mild correlation
in variable gamma-ray sources between the variability and spectral
indices, as seemingly observed in Figure~4. See the work by Punsly
et al. (2000) for details on the model.

Apart from 3EG J1828+0142, the source 3EG J0241+6103 (2EG
J0241+6119) was already suggested as a possible identification of
a magnetized black hole by one of us (Punsly 1999a,b). The source
2EG J0241+6119 is known to be variable since the work by Tavani et
al. (1998), who discarded a pulsar origin. This source appeared in
a search for gamma-ray emitters, with strong emission also in
X-rays, but weak output in radio. The binary system LSI
+61$^o$303, which is within the 95\% probability contour of the
gamma-ray detection, contains an unseen compact object suggested
as the possible generator of the observed gamma rays (Punsly
1999a, b). However, a positive identification with the 3EG
detection is not conclusive because of the fact that there is no
signature in the gamma emission that can be correlated with the
binary dynamics. Note that the magnetized black hole model in
Punsly (1999b) does not produce any modulation of the gamma-ray
signal on orbital time scales to a first order.

\section{Conclusions}

In this paper we have shown that there exists a group of
unidentified, low-latitude, gamma-ray sources with the following
characteristics:
\begin{enumerate}
  \item They do not positionally coincide with any known galactic
  object such as Wolf-Rayet or Of massive stars, catalogued SNRs,
  and OB associations. The probability that a set of 40 sources,
  with the same positional errors as the EGRET ones,
  lacks any positional correlation with these galactic objects
  totally by chance is less than  $3 \times 10^{-4}$.
  \item Even considering positional coincidences with 101
  identified candidates to SNRs, in addition to Green's (1998) catalog,
  only 7 out of 40 sources present correlation, which is consistent with
  being the mere effect of chance.
  \item The sample presents interesting variability features: 32 \% are variable and
  45 \% of them are dubious cases.
  \item The sample presents
  a steep average spectral index $<\Gamma> = 2.41 \pm 0.2$; only 3 out of
  40 sources have $\Gamma < 2$.
  \item Due to isotropy constraints on the average density
  of active galactic nuclei, not all of
  the 40 sources considered here can be AGNs
  seen through the disk of galaxy.
  The surface density of  variable unidentified sources at low
  galactic latitudes is five times higher than that of AGNs out of the plane.
  This, and the fact that there are 5 AGNs already detected
  at $| b | \leq 10\degr$, makes that
  if the distribution
  of AGNs is isotropic, just a few out of 40 gamma-ray sources could be AGNs.
  Indeed, if we extrapolate the AGN number density out of the
  plane to latitudes within $(-10\degr,10\degr)$, we would expect
  6.5 AGNs to be detected by EGRET, leaving room for just a couple
  of remaining unidentified objects to be AGNs, one, perhaps, being
  B2013+370 (Mukherjee et al. 2000).

\end{enumerate}

Altogether, these features are not in accordance with usual models
of gamma-ray emitters: they cannot be explained by pulsars, SNRs
in interaction with nearby clouds, massive stars, or AGNs. In
addition, we have found an apparent trend for sources with a
higher degree of variability to present steeper photon spectral
indices. This latter result, however, needs confirmation with the
more complete sample to be obtained by the next generation of
gamma-ray satellites (Gehrels et al. 1999). With the currently
large positional error boxes, we have no hope to completely
identify the source for all these detections, but if, after the
launch of future gamma-ray satellites the previously quoted
characteristics are confirmed, a new population of objects
probably will be needed.

To our knowledge, the only model capable of reproducing all the
peculiar features presented by some of the sources in our sample
(steep spectral gamma-ray index, high variability, absence of
clear lower-frequency counterparts, etc) is that of isolated
Kerr-Newman black holes (Punsly et al. 2000). This does not
exclude the possibility that alternative models based on, for
instance, stellar high-energy variability (Benaglia et al. 2000),
or pulsar abnormal activity (e.g. Tavani and Arons 1997), could
play an important role explaining some detections.

The hypothesis of a population of magnetized black holes in the
Galaxy can be tested with the forthcoming GLAST and INTEGRAL
satellites, which will allow investigators to increase the sample
and measure the spectral energy distribution for particular
sources with high precision. In the case of INTEGRAL, the
spectrometer SPI and the imager IBIS  could detect the
electron-positron annihilation features of the black hole jets, as
indicated by Punsly et al. (2000).

\begin{acknowledgements} This work was partially supported by CONICET
(D.F.T., G.E.R., J.A.C., and P.B.), ANPCT (PICT 98 No. 03-04881),
and by Fundaci\'{o}n Antorchas, through separate grants to D.F.T.,
G.E.R., and J.A.C. H.A. thanks CONACyT (Mexico) for financial
support under grant 27602-E. This research made use of the
NASA/IPAC Extragalactic Database (NED) which is operated by the
Jet Propulsion Laboratory, California Institute of Technology,
under contract with the National Aeronautics and Space
Administration. Tables are also available on-line at
http://www.iar.unlp.edu.ar/garra. We gratefully acknowledge
criticism and comments from an anonymous referee, which resulted
in a substantial improvement of the paper.
\end{acknowledgements}

\section*{Appendix A: SNR candidates}

In this appendix we present the complete list of 101 candidates to
SNRs that were used in our study of positional coincidences. This
list is presented electronically in Table~6 and it consists of
three columns: galactic coordinates $(l,b)$ and the diameter of
the extended source ($\theta$) in arcminutes. The positions of
these sources were reported by Duncan et al. (1995), Combi \&
Romero (1998), Combi et al. (1998b, 1999a), and Jonas (1999). In
general, these new candidates are much more extended and weaker
than those previously known.

\section*{Appendix B: Variability of AGNs}

In this appendix we present data for the variability indices of
the 66 AGNs already detected in the 3EG catalog. These data were
used both in Figure~4 and for making a comparison with our sample
in the main text. We provide electronically (Table~7) the same
columns as those given in Table~1 for the unidentified sources.

\section*{Appendix C: Variability analysis of all low latitude
gamma-ray sources}

If we combine the present work with that of Zhang et al. (2000),
(taking into account the same normalization for $<\mu>_{{\rm
pulsars}}$, as explained above) who studied the variability of the
sources which were positionally coincident with OB associations
and SNRs as discovered by Romero et al. (1999), we can now make an
assessment of the variability of the whole sample of unidentified
sources at low galactic latitudes. There are 81 such sources, six
of them (3EG J0824$-$4610, 0827$-$4247, 0828$-$4954, 0841$-$4356,
0859$-$4257 and 0848$-$4429) were recently reported to be
artifacts produced by the high intensity emission of the nearby
Vela pulsar. Another one, 3EG J0747$-$3412, was found to be
coincident only with a WR star by Romero et al. (1999), and its
variability index is here reported for the first time to be
$I=2.81$. For future reference we present electronically in
Table~8 the 3EG source name and the variability index $I$ for all
low-latitude unidentified sources, normalized to the gamma-ray
pulsars detected in the 3EG catalog.

\newpage

\renewcommand{\baselinestretch}{1.4}
\begin{table*}
\caption[]{Unidentified 3EG sources without known possible
galactic counterparts.}
\begin{flushleft}
\begin{tabular}{l c r c c c c c c c}
\hline \noalign{\smallskip} $\gamma$-Source &  $l$ & $b$ &
$\Delta\theta$ & $\Gamma$ & $\left< F_{\gamma}\right> \times
10^{-8}$ & $\sigma_{{\rm sd}}$ & $N_{{\rm vp}}$ & $\mu$ & $I $\cr
(3EG J)& (deg) & (deg) & & &(ph cm$^{-2}$ s$^{-1}$) & & & & \\
\noalign{\smallskip} \hline \noalign{\smallskip}
 0241 $+$6103$^\star$ & 135.87 &    0.99 &   0.18 & $2.21 \pm 0.07$ & 69.4
 &  24.6 &9 & 35.4     &1.31  \\
 0323 $+$5122  & 145.64 &  $-$4.67 &   0.55 & $2.38 \pm 0.41$
 & 17.5 &9.5 & 6& 54.2      &2.01 \\
0416 $+$3650$^\dag$ & 162.22  & $-$9.97 &   0.63 & $2.59 \pm 0.32$
&21.1 & 14.9&10 & 70.6 &2.61 \\
0435 $+$6137 & 146.50 &   9.50 &   0.66  &$2.46 \pm 0.35$ &18.8
& 5.1 & 4& 27.1          &1.00  \\
0459 $+$3352 &  170.30 &  $-$5.38 &   0.98 &$2.54 \pm 0.24$ &21.9
&  7.6 & 11& 34.7       &1.28 \\
0500 $+$2529 & 177.18 & $-$10.28 &   0.36 &$2.52 \pm 0.32$ &14.2
& 8.4 & 17& 59.1        &2.19 \\
 0510 $+$5545 & 153.99  &  9.42 &   0.71 &$2.19 \pm 0.20$ &22.5
 &  11.3 &9 & 50.2        &1.86 \\
0520 $+$2556 & 179.65 &  $-$6.40 &   0.86 &$2.83 \pm 0.24$ &25.0
&  8.5 &13 & 30.0       &1.11 \\
0521 $+$2147 & 183.08 &   $-$8.43  &  0.45 &$2.48 \pm 0.15$ &28.5
&  13.9 & 15& 48.8     &1.81 \\
0533 $+$4751 & 162.61 &   7.95  &  0.60 &$2.55 \pm 0.23$ &18.2
&  6.1 & 8& 33.5 &         1.24 \\
0546 $+$3948 & 170.75 &   5.74 &   0.67&$2.85 \pm 0.21$ &17.5
& 8.1 & 13& 46.3 &          1.71 \\
0556 $+$0409& 202.81  &$-$10.29 &   0.47&$2.45 \pm 0.16$ &19.3 &
8.8
& 10& 45.6          &1.69 \\
 0628 $+$1847 & 193.66 &   3.64 &    0.57 &$2.30 \pm 0.10$ &32.9
  &12.0 & 10& 36.5        &1.35 \\
0903 $-$3531 & 259.40 &    7.40 &   0.58 &$2.66 \pm 0.24$ &22.4
& 9.4 &6 & 41.9          &1.55 \\
1014 $-$5705 & 282.80 &  $-$0.51 &    0.67 &$2.23 \pm 0.20$ &47.5
& 18.9 &10 & 39.8      &1.47 \\
1316 $-$5244 & 306.85 &   9.93 &    0.50 &$2.54 \pm 0.18$
&19.7 &11.8&11 & 59.8          &2.21 \\
1631 $-$4033 & 341.61  &  5.24 &   0.89 &$2.25 \pm 0.27$ &32.3
 &11.0 & 11& 34.0          &1.26 \\
1633 $-$3216&348.10 &  10.48  &  0.87 &$2.58 \pm 0.24$ &16.1 &
9.5&12 & 59.0
&            2.18 \\
 1638 $-$5155 & 334.05 &  $-$3.34 &   0.68 &$2.56 \pm 0.21$ &45.3
  & 29.9& 15& 66.0       &2.44 \\
1704 $-$4732 & 340.10 &   $-$3.79 &   0.66 &$1.86 \pm 0.33$ &
29.5& 23.6&18  & 80.0       &2.96 \\
1717 $-$2737 & 357.67 &   5.95 &   0.64 &$2.23 \pm 0.15$
&32.9 &17.8 & 18& 54.1          &2.00 \\
1735 $-$1500 & 10.73 &    9.22 &   0.77 &$3.24 \pm 0.47$
&19.0 &45.4 & 19& 238.9         &8.86 \\
1736 $-$2908 & 358.79 &   1.56 &   0.62 &$2.18 \pm 0.12$
 &49.6 &32.4 & 21& 65.3          &2.42 \\
 1741 $-$2050 & 6.44 &     5.00  &  0.63 &$2.25 \pm 0.12$
  &33.4 & 19.1& 19& 57.2         &2.12 \\
1741 $-$2312 & 4.42 &     3.76 &   0.57 &$2.49 \pm 0.14$
&34.0 &19.8 & 19& 58.2          &2.15 \\
 1744 $-$3934& 350.81 &  $-$5.38&    0.66  &$2.42 \pm 0.17$
 &26.5 & 21.8& 20& 82.2       &3.04 \\
1746 $-$1001& 16.34 &    9.64 &    0.76 &$2.55 \pm 0.18$
&29.7 &25.6 & 18& 86.2          &3.19 \\
1757 $-$0711& 20.30  &   8.47  &  0.68 &$2.51 \pm 0.20$
 &33.6 &18.2 & 15& 54.2           &2.01 \\
1800 $-$0146 & 25.49  &  10.39 &   0.77&$2.79 \pm 0.22$
&25.5 &13.3 & 14& 52.1           &1.93 \\
1810 $-$1032 & 18.81  &   4.23 &   0.39&$2.29 \pm 0.16$
 &31.5 &22.2 & 17& 70.5           &2.61 \\
 1812 $-$1316&16.70 &    2.39  &  0.39&$2.29 \pm 0.11$
 &43.0 & 30.2& 18& 70.3            &2.60 \\
1828 $+$0142& 31.90 &    5.78 &   0.55&$2.76 \pm 0.39$
&30.8 &44.3 & 8& 143.8            &5.33 \\
1834 $-$2803&5.92 &    $-$8.97  &  0.52&$2.62 \pm 0.20$
&17.9 &13.7 &20 & 76.5           &2.83 \\
 1837 $-$0606&25.86  &   0.40  &  0.19&$1.82 \pm 0.14$
  &57.5 & 37.4& 12& 65.0            &2.41 \\
1904 $-$1124 & 24.22 &   $-$8.12  &  0.50&$2.60 \pm 0.21$
&22.5 &17.7 & 14& 78.6         &2.91 \\
1928 $+$1733& 52.91  &   0.07  &  0.75&$2.23 \pm 0.32$
&38.6 &41.6 & 10& 107.7           &3.99 \\
1958 $+$2909&66.23 &   $-$0.16  &  0.57&$1.85 \pm 0.20$
&35.6 &16.2 &10 & 45.5           &1.68 \\
2035 $+$4441&83.17 &    2.50 &   0.54&$2.08 \pm 0.26$
&39.1 & 35.4&14 & 90.5             &3.35 \\
 2100 $+$6012$^\dag$& 97.76  &   9.16 &    0.48&$2.21 \pm 0.25$
  &23.3 & 8.7& 7& 37.3     &1.38 \\
2206 $+$6602$^\dag$& 107.23 &   8.34  &  0.88 &$2.29 \pm 0.26$&
25.4 & 4.6 & 4& 18.1      &0.67 \\
\hline
\multicolumn{9}{l} {$\star$: This source also displays short-term
variability (Wallace et al. 2000).}\cr
\end{tabular}
\end{flushleft}
\end{table*}

\begin{table*}
\caption[]{Statistical variability of pulsars. The last one was
identified by Cusumano et al. (2000). }
\begin{flushleft}
\begin{tabular}{l l c r c c c c c }
\hline \noalign{\smallskip} $\gamma$-Source & Pulsar Name & $l$ &
$b$ & $\left< \Gamma \right>$ & $\left< F_{\gamma}\right> \times
10^{8}$ & $\sigma_{{\rm sd}}$ & $N_{{\rm vp}}$ & $\mu$  \cr (3EG
J)& &(deg) & (deg) & &(ph cm$^{-2}$ s$^{-1}$) & &  \\
\noalign{\smallskip} \hline \noalign{\smallskip}

 0534 $+$2200 & Crab & 184.5 & $-$5.8 & 2.19 &  219.0 & 29.2 &  16 & 13.3 \\

 0633 $+$1751 & Geminga & 195.0 & 4.3 & 1.66 &  350.9 & 50.1 &  14 & 14.3\\

 0834 $-$4511 & Vela & 263.5 & $-$2.8 & 1.69 &  848.2 & 183.6 &  8 & 21.6 \\

 1058 $-$5234 & PSR B1055$-$52 & 286.1 & 6.5 & 1.94 &  36.1 & 14.6 &  15 & 40.4 \\

 1710 $-$4439 & PSR B1706$-$44 & 343.0 & $-$2.8 & 1.86 &  109.9 & 44.5 &  20 & 40.5 \\

 0634 $+$0521 &  & 206.1 & $-$1.4 & 2.03 &  25.5 & 7.2 &  9 & 27.6 \\


\hline
\end{tabular}
\end{flushleft}
\end{table*}


\begin{table*}
\caption[]{Positional coincidences with candidates to SNRs.}
\begin{flushleft}
\begin{tabular}{l c r c c }
\hline \noalign{\smallskip} $\gamma$-Source & $l$ & $b$ & $\delta$
& $\Delta$ \cr (3EG J)& (deg) & (deg)& (deg) & (deg)
\\ \noalign{\smallskip} \hline \noalign{\smallskip}

  0903 $-$3531 &   260.20  &   1.40  & 6.05 &  13.58 \\

  1631 $-$4033 &   342.60  &   8.20 &  3.11 &  3.89 \\

 1638 $-$5155 &   333.00 &    0.00 & 3.50 & 3.68 \\

 1704 $-$4732 &   340.80 &    $-$4.80 & 1.22 &  2.11 \\

 1717 $-$2737  &   356.90  &   8.50 & 2.66 & 5.39 \\

 1834 $-$2803   &  7.30 &   $-$5.30 & 3.91 & 4.12 \\

 1837  $-$0606  &  27.00   &  0.50 & 1.14 & 3.69 \\

\hline
\end{tabular}
\end{flushleft}
\end{table*}

\begin{table*}
\caption[]{Characteristics of the point-like radio sources inside
the $\gamma$-ray contour of 3EG J1631-4033}
\begin{tabular}{ c c c c c c c c}
\hline Source & ($l$, $b$) & $F_{\rm 2.4 GHz}$ & $F_{\rm 4.8
GHz}$& $F_{\rm 8.3 GHz}$  & $\alpha_{2.4/4.8}$  &
$\alpha_{4.8/8.3}$ & ID \\
       & (deg, deg) &      (mJy)        &      (mJy)       &   (mJy)            &                     &          &   \\
\hline S1     & (341.6,$+6.2$) &  877.3  &  670.1 &  --   & $-$0.3
& --   & PMN J1631$-$4015 \\ S2     & (341.8,$+5.4$) &  892.1  &
490.0 & 150.2 & $-$0.8  & $-$2.0 & PMN J1627$-$3952 \\ S3     &
(341.9,$+4.1$) & 2260.3  & 1120.0 & 700.0 & $-$1.0  & $-$0.8 & PMN
J1636$-$4101 \\ S4     & (342.1,$+5.6$) &  520.2  &  180.0 &  50.0
& $-$1.5  & $-$2.2 & PMN J1631$-$3956 \\ \hline
\end{tabular}
\end{table*}

\begin{figure}
\resizebox{\hsize}{!} {\includegraphics{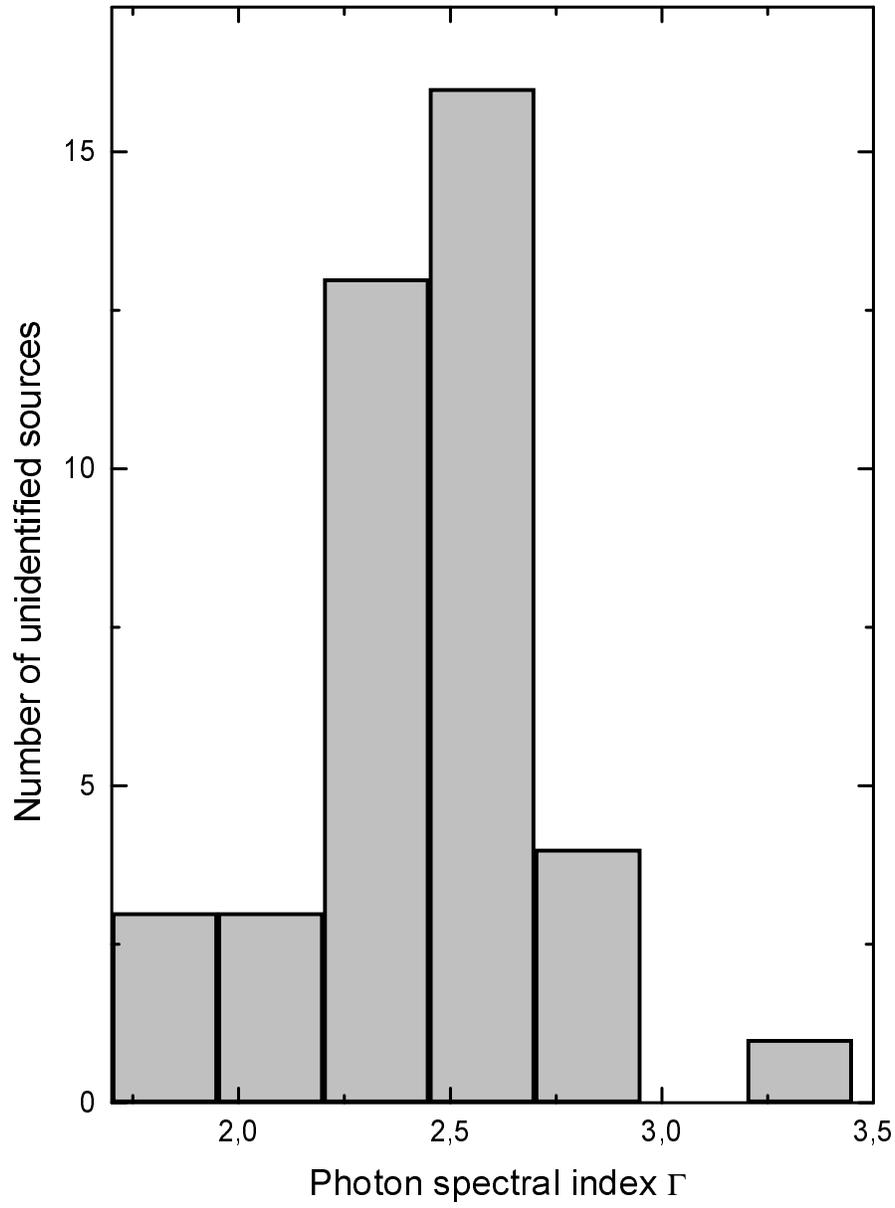}}
\caption{Distribution of the spectral indices of the 40
unidentified sources.} \label{fig1}
\end{figure}

\begin{figure}
\resizebox{\hsize}{!}{\includegraphics{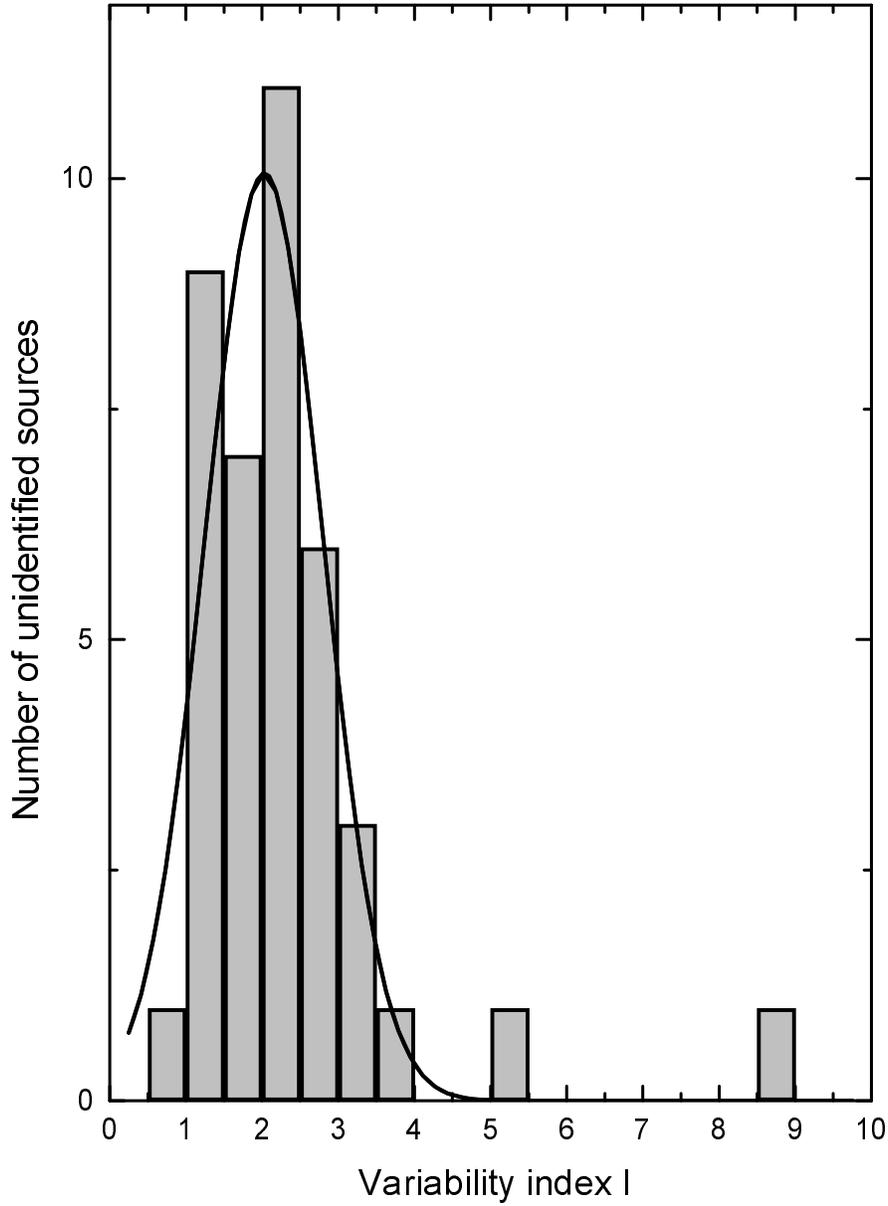}}
\caption{Variability index distribution. The spline curve is the
best Gaussian fit, it peaks at $I=$2.0 and presents a standard
deviation of $\sigma_I=$0.7. These values suggest that many of the
unidentified sources are variable.}\label{fig2}
\end{figure}

\begin{figure}
\resizebox{\hsize}{!}{\includegraphics{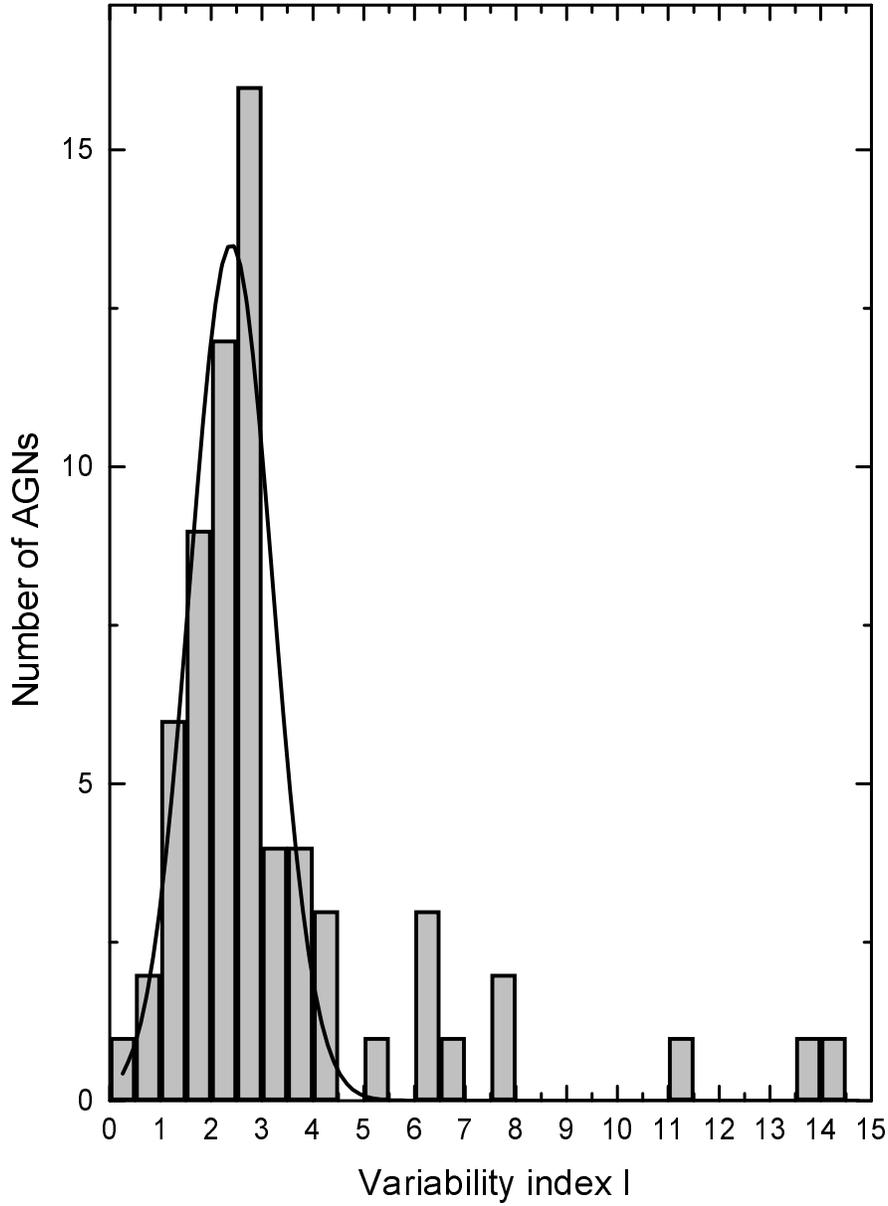}}
\caption{Variability index distribution for the 66 AGNs in the 3EG
catalog. The spline curve is the best Gaussian fit, it peaks at
$I=2.3$ and has a standard deviation of
$\sigma_I=$0.8.}\label{fig3}
\end{figure}

\begin{figure}
\resizebox{\hsize}{!}{\includegraphics{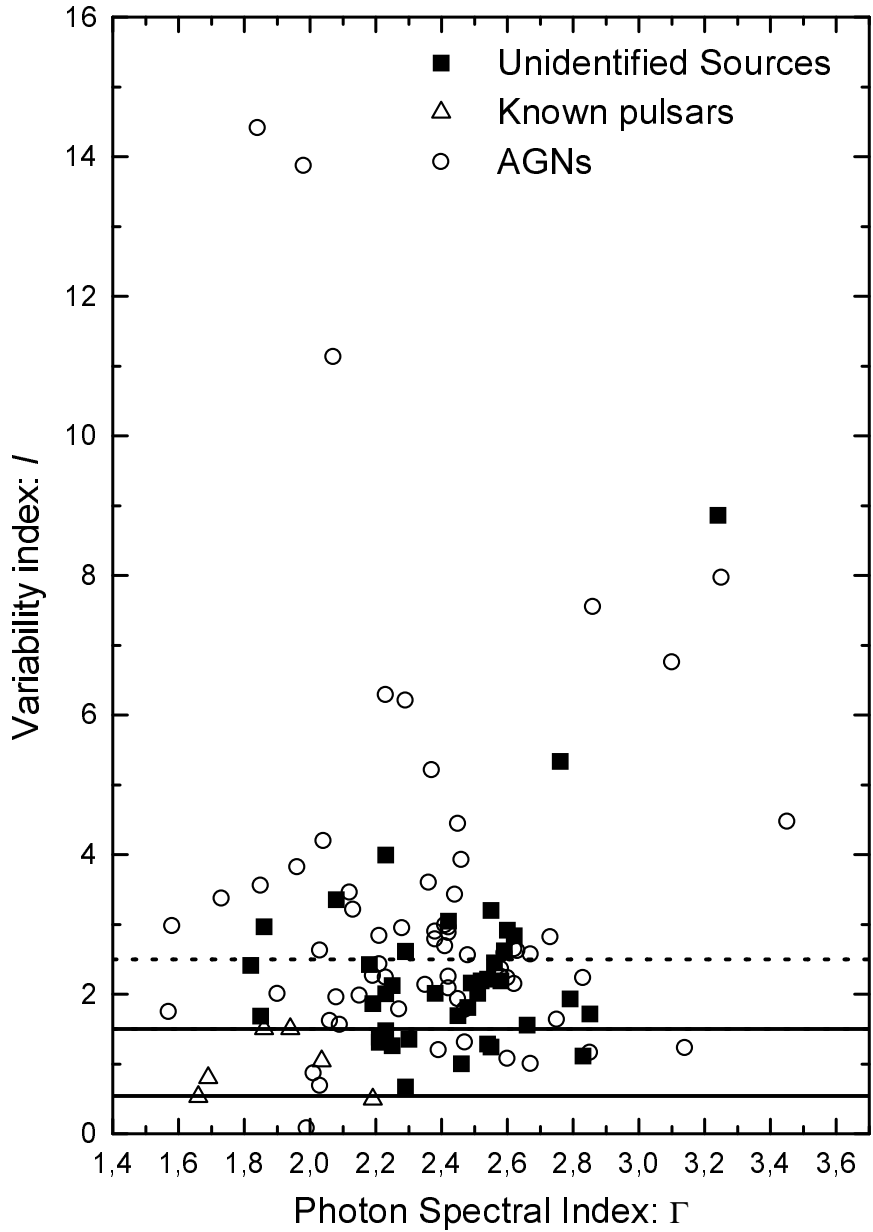}}
\caption{Variability versus photon spectral index plot. We also
present the values for pulsars and 66 AGNs in the 3EG catalog.
Horizontal solid lines represent the lowest and maximum values for
pulsars. The dotted line corresponds to the limit above which all
sources are variable (3$\sigma$).}\label{fig4}
\end{figure}

\begin{figure}
\resizebox{\hsize}{!}{\includegraphics{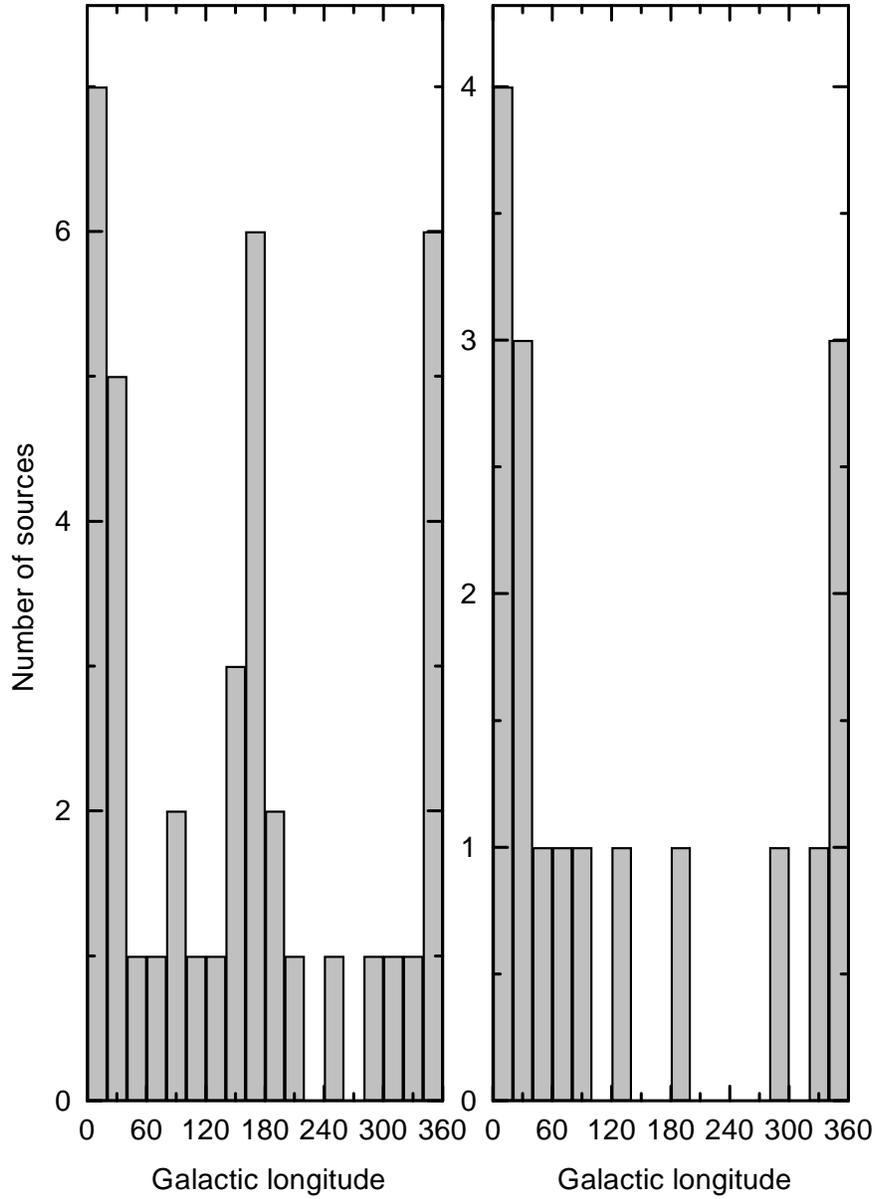}}
\caption{Galactic longitude distribution of the 3EG sources
considered in our sample, for different flux lower limits. See
text for explanation.}\label{lon}
\end{figure}

\begin{figure*}
\resizebox{8cm}{!} {\includegraphics{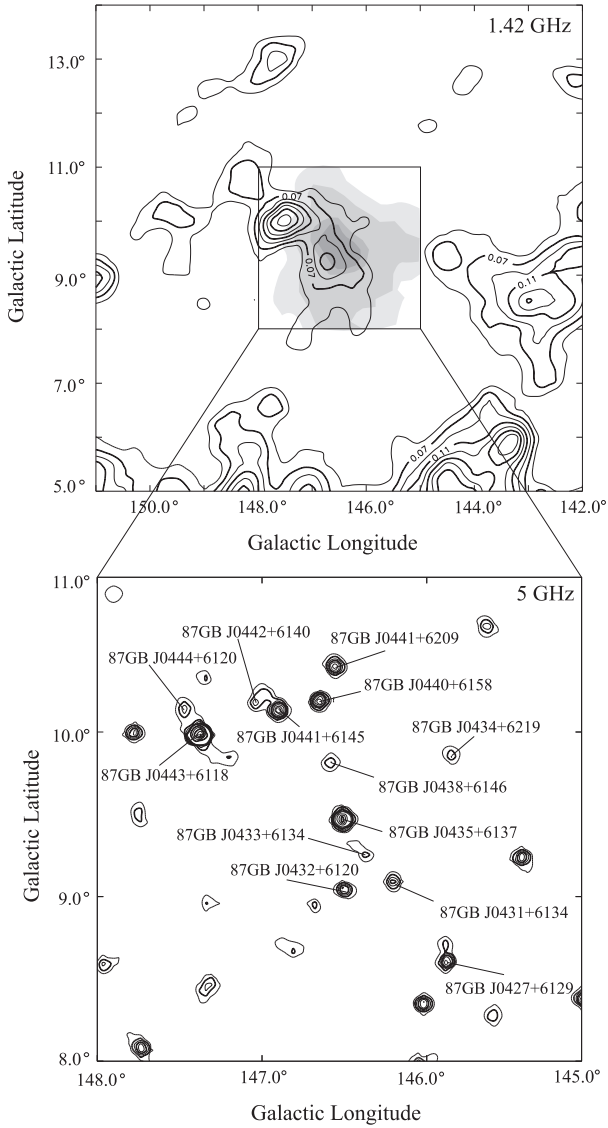}}\hfill
\parbox[b] {50mm}{\caption{{\bf Upper panel:} Map of the radio
field around 3EG J0435+6137 at 1.42 GHz (angular resolution of
34') with the diffuse emission filtered using a beamwidth of 120'
$\times$ 120'. Contours are labeled in steps of 0.02 K from 0.05
to 0.22 K in brightness temperature. The superimposed grey-scaled
levels represent the 99\%, 95\%, 68\%, and 50\% statistical
probability that the gamma-ray source lies within each contour.
{\bf Lower panel:} Image at 5 GHz of the central field obtained
with the former NRAO 91-m telescope at Green Bank (Condon et al.
1994). Contours are shown starting at 10 mJy beam$^{-1}$, in steps
of 10 mJy beam$^{-1}$.}\label{fig5} }
\end{figure*}

\begin{figure*}
\resizebox{8cm}{!} {\includegraphics{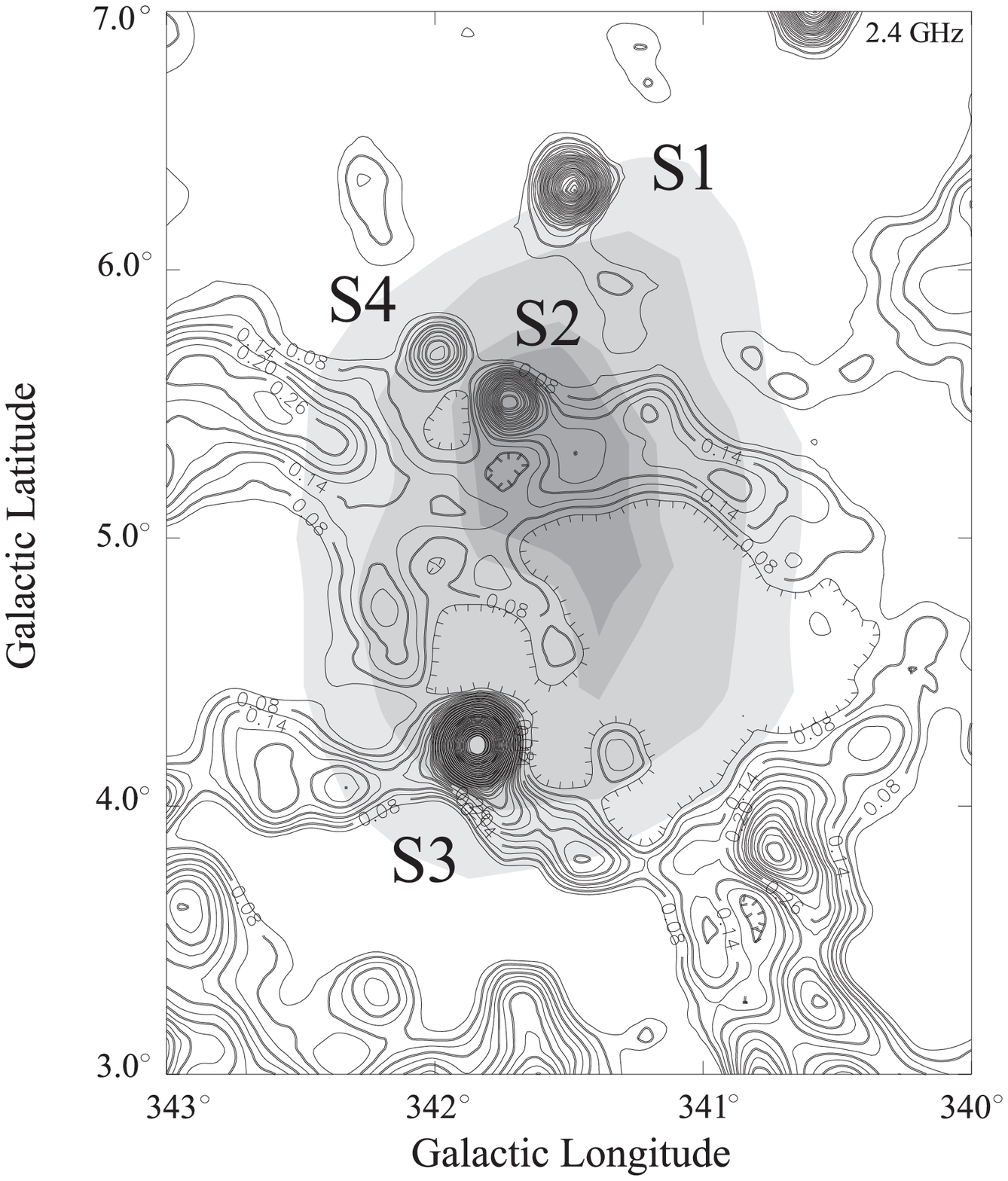}} \hfill
\parbox[b] {50mm}{\caption{Radio image of the field of 3EG
J1631-40 at 2.4 GHz (angular resolution of 10.4') after the
subtraction of the diffuse emission. Superimposed on the radio
image we show the confidence contours of the likelihood test
statistics of the EGRET detection. The best result for revealing a
counterpart of the gamma-ray source is obtained for a filtering
beamwidth of 90' $\times$ 90' and 6 iterations. Radio contours are
shown at steps of 0.05 from 0.05 to 6 Jy
beam$^{-1}$.}\label{fig6}}
\end{figure*}


\begin{thebibliography}{}

\bibitem{} Benaglia P., Romero G.E., Stevens I. and Torres D.F.,
A\&A, to appear [astro-ph/0010605]
\bibitem{} Cass\'e M., Paul J.A. 1980, ApJ 237, 236
\bibitem{} Combi J.A., Romero G.E. 1995, A\&A 303, 873
\bibitem{} Combi J.A., Romero G.E. 1998, A\&AS 128, 423
\bibitem{} Combi J.A., Romero G.E., Benaglia P. 1998a, A\&A 333, L91
\bibitem{} Combi J.A., Romero G.E., Arnal, M. 1998b, A\&A 333, 298
\bibitem{} Combi J.A., Romero G.E., Benaglia P. 1999a, AJ 118, 659
\bibitem{} Combi J.A., Romero G.E., Benaglia P. 1999b, ApJ 519, L177
\bibitem{} Condon J.J., Broderick J.J., Seielstad G.A. 1989, AJ
97, 1064
\bibitem{} Condon J.J., Broderick J.J.,
Seielstad G.A., Douglas K., Gregory,P.C. 1994,  AJ 107, 1829
\bibitem{} Cusumano G., Maccarone M.C.,
 Nicastro L., Sacco B., Kaaret P. 2000, ApJ 528, L25
\bibitem{} Duncan A.R., et al. 1995, MNRAS 277, 36
\bibitem{} Esposito J.A., Hunter S.D., Kanbach G., Sreekumar P. 1996,
ApJ 461, 820
\bibitem{} Gehrels N., Michelson B., 1999, Astroparticle Phys.,
11, 277
\bibitem{} Gehrels N., Macomb, D. J., Bertsch D. L., Thompson D. J.,
Hartman R. C., 2000, Nature, 404, 363
\bibitem{} Griffith M.R., Wright A.E. 1993, AJ 105, 1666
\bibitem{} Green D.A. 1998, A Catalogue of Galactic Supernova Remnants,
Mullard Radio Astronomy Observatory, Cambridge, UK (available at
http://www.mrao.cam.ac.uk/surveys/snrs/)
\bibitem{} Grenier I. A. 1995, Adv. Space. Res. 15, 73
\bibitem{} Hartman R.C., Bertsch D.L., Bloom S.D., et al. 1999, ApJS 123, 79
\bibitem{} Hartman R.C. 2000, private communication.
\bibitem{} Jonas J. 1999, Ph.D. Thesis, Rhodes University,
South Africa.
\bibitem{} Kaspi V.M., Lackey J.R., Mattox J.,
 Manchester R. N., Bailes M., Pace R. 2000, ApJ 528, 445
\bibitem{} McLaughlin M.A., Mattox J.R., Cordes J.M., Thompson D.J. 1996,
ApJ 473, 763
\bibitem{} Mirabel I.F., Rodr\'{\i}guez L.F., 1998, Nature, 392,
673
\bibitem{} Montmerle T. 1979, ApJ 231, 95
\bibitem{} Mukherjee R., Grenier I.A., Thompson D.J. 1997, Proceedings of
the Fourth Compton Symposium, C. D. Dermer, M. S. Strickman and J.
D. Kurfess Eds., AIP, New York, p.384
\bibitem{} Mukherjee R., Gotthelf E.V., Halpern J., Tavani M. 2000, ApJ to appear
 [astro-ph/0005491]
\bibitem{} Nice D., Sawyer T., 1997, ApJ, 476, 261
\bibitem{}\"{O}zel M.E., Thompson D.J., 1996, ApJ, 463,
\bibitem{} Punsly B. 1998a, ApJ 498, 640
\bibitem{} Punsly B. 1998b, ApJ 498, 660
\bibitem{} Punsly B. 1999a, ApJ 516, 141
\bibitem{} Punsly B. 1999b, ApJ 519, 336
\bibitem{} Punsly B., Romero G.E., Torres D.F., Combi J.A 2000,
A\&A, 364, 552.
\bibitem{} Reich W., Reich P. 1986, A\&AS 63, 205
\bibitem{} Romero G.E., Combi J.A., Colomb F.R. 1994, A\&A 288,
731
\bibitem{} Romero G.E., Benaglia P., Torres D.F. 1999 A\&A 348,
868 (Paper I).
\bibitem{} Sturner S.J., Dermer C.D. 1995, A\&A 293, L17
\bibitem{} Tavani M., Mukherjee R., Mattox J.R., et al. 1997,
ApJ 479, L109
\bibitem{} Tavani M., Arons J. 1997, ApJ 477, 439
\bibitem{} Tavani M., Kniffen D.,
 Mattox J. R., Paredes, J. M.,
 Foster R. 1998, ApJ 497, L89
\bibitem{} Thompson D.J., Arzoumanian Z.,
 Bertsch D.L., et al. 1994, ApJ 436, 229
\bibitem{} Thompson D.J., Bertsch D.L.,
Dingus B.L., et al. 1995, ApJS 101, 259
\bibitem{} Thompson D.J., Bertsch D.L.,
Dingus B.L., et al. 1996, ApJS 107, 227
\bibitem{} Tompkins W. 1999, Ph.D. Thesis, Stanford University.
\bibitem{}Verkhodanov O.V., Trushkin S.A., Andernach H., Chernenkov V.N.: 1997,
Astronomical Data Analysis Software and Systems -- VI,
ASP~Conf.\,Ser. 125, 322, eds.~G.\,Hunt \& H.E.\,Payne,
[astro-ph/9610262]
\bibitem{} Wallace P.M., Griffis N.J., Bertsch D.L., et al.
2000, ApJ 540, 184
\bibitem{} Yadigaroglu I.-A., Romani R.W. 1997, ApJ 476, 356
\bibitem{} Zhang L., Zhang Y.J., Cheng K.S. 2000, A\&A, 357, 957

\end{thebibliography}
\end{document}